# Characteristics of mono-, di-, and trivalent cations in electric double layers: a molecular dynamic investigation

Bowen Ai,[1,2] Zekun Gong,[1,2] Long Ma,[1,2] Hongwen Zhang,[1,2] Tianyi Sui,[3]* and Yinghua Qiu[1,2]*

1. Key Laboratory of High Efficiency and Clean Mechanical Manufacture of the Ministry of Education, State Key Laboratory of Advanced Equipment and Technology for Metal Forming, School of Mechanical Engineering, Shandong University, Jinan, 250061, China

2. Shenzhen Research Institute of Shandong University, Shenzhen, 518000, China

3. School of Mechanical Engineering, Tianjin University, Tianjin, 300072, China

*Corresponding author: yinghua.qiu@sdu.edu.cn; suity@tju.edu.cn

**Abstract**

Ionic behaviors, including ion distributions and hydration characteristics at solid-liquid interfaces, are important research interests in many important applications, such as electric double-layer capacitors and water lubrication. Here, we systematically investigated the concentration distributions, hydration numbers, and screening properties of $Li^+$, $Na^+$, $K^+$, $Ca^{2+}$, $Mg^{2+}$, and $La^{3+}$ ions inside electric double layers (EDLs) at various charge densities (σ). For σ weaker than −0.16 C/m$^2$, monovalent cations mainly accumulate in the outer Helmholtz plane (OHP). As σ magnitude increases, monovalent cations start to dehydrate and migrate to the inner Helmholtz plane (IHP), following the order of $K^+$, $Na^+$, and $Li^+$. This size-dependent behavior arises from the lower hydration energy of larger ions. While for the di- and trivalent ions, no obvious IHP appears. Based on ion distributions, the screening effect of counterions on surface charges is evaluated by analyzing the net charge distributions. As σ changes from 0 to −0.32 C/m$^2$, due to the stronger accumulation of cations in EDLs, the location of the neutral plane changes from ~12 to ~4 Å. When σ reaches a threshold, excessive accumulation of cations can induce charge inversion. The threshold value and the maximum reversed charge are found to correlate with the ion size, cation valence, and concentration.

## Introduction

At solid-liquid interfaces, solid surfaces can take charges due to electrical charging, protonation of surface groups, or ion adsorption.[1] Under electrostatic interactions, counterions can be attracted to the charged surface to form electric double layers (EDLs), which typically consist of the Stern layer and diffusion layer. Inside EDLs, the distribution of counterions is of great importance in many practical applications, including the electric double-layer capacitor,[2] ion conductance,[3] energy conversion,[4-5] and water lubrication.[6] Therefore, it is crucial to understand the ion distribution inside EDLs, which can shed light on the fundamental properties of EDLs at solid-liquid interfaces.[7-8]

The Poisson-Boltzmann (PB) equations[1, 9] provide the theoretical description of the ion distribution inside the diffusion layer. However, as a mean-field theory, they do not account for ion size, hydration properties, or the discreteness of water molecules.[1, 10] Near negatively charged surfaces, the monovalent counterions such as $Li^+$, $Na^+$, or $K^+$ ions, share the same ion distribution predicted by the PB equations, despite their different ion sizes and hydration properties. Also, the PB equations are mainly suitable for the prediction of the ion distribution in dilute and monovalent solutions at a low surface charge density. This theory fails under high surface charge densities, in concentrated solutions, or with multivalent counterions, where the correlation effect among surface charges and ions becomes significant.[11] The Stern layer, adjacent to the charged surface, includes the inner (IHP) and outer Helmholtz planes (OHP), whose distribution of ions falls outside the prediction of the mean-field theory.

Monovalent solutions such as LiCl, NaCl, and KCl are widely applied. For example, the feed solution in seawater desalination is primarily 0.5 M NaCl.[12] The ion separation process typically aims to achieve the efficient separation of $Li^+$ ions from $Mg^{2+}$ ions in salt lake brines.[13] The concentration distribution and hydration state of ions affect the efficiency of membrane-based processes for seawater desalination.[14-15] For the detection of biomolecules, NaCl and KCl solutions with a physiological concentration are usually considered.[16] The distribution and hydration state of counterions also affect the recognition accuracy of biomolecular structures.[17] From the previous research, the ion size and valence have significant impacts on the ion distribution and solvation structure.[18-20] Recent studies have gradually revealed the complexity of ion behaviors inside EDLs.[21] Using MD simulations, Zhou et al. investigated the hydration behavior in infinitely dilute aqueous solutions at 298 K. They revealed the relationship between ionic radius and hydration strength, establishing a cation hydration strength order of $Li^+$ > $Na^+$ > $K^+$.[22] Through machine learning-based potential function simulations, Zhang et al.[23] found that there is a continuous exchange of ions in the Stern layer at the $TiO_2$-liquid interface, where the ion residence time in IHP is at the nanosecond level, significantly longer than the picosecond scale of the intermediate Helmholtz layer, indicating that there are kinetic differences between the IHP and OHP regions. Near charged surfaces, the charge shielding efficiency depends mainly on the valence rather than the size of counterions.[24] For the case of overscreening, charge inversion appears as a counterintuitive phenomenon.[11] Based on the surface force measurement with AFM,

Besteman et al.,[25] confirmed the ion valence as the dominant factor for charge inversion.

The structure and behavior of ions at the solid-liquid interface have been extensively probed through both experimental and simulation approaches. Through in situ high-resolution X-ray reflection and resonant anomalous X-ray reflection experiments, Carr et al.[26] studied the "over-adsorption" phenomenon of trivalent $Y^{3+}$ ions on the negatively charged graphene surface, where the number of adsorbed ions far exceeded the amount needed to neutralize the surface charges. Using molecular dynamics simulations with a polarizable force field, Pykal et al.[27] explored the ion distribution at the interface between graphene and aqueous solutions of KF and KI. They found that $F^-$ ions were selectively excluded from the graphene surface, while $I^-$ ions presented a strong affinity to the graphene layer. These studies have explored related interfacial phenomena, including polarization-driven adsorption[28], hydration structure of the ions[29], solvent structuring[30], and ion-selectivity in nanopore graphene[31]. Collectively, however, these previous investigations have predominantly centered on ion adsorption phenomena at the solution-graphene interface, with comparatively less emphasis on the detailed distribution of ions within the entire EDLs.

Molecular dynamics (MD) simulations can provide an effective tool for the investigation of the nanoscale physical properties at solid-liquid interfaces, including the ion distribution and hydration characteristics,[32-33] which are currently challenging to be directly explored with experiments due to the limited technical conditions.

Although many explorations of the microscale characteristics at solid-liquid interfaces have been carried out with MD simulations,[34] and provided significant insights into the ion structuring[35], adsorption trends[36], and ion-surface interactions at the graphene-electrolyte interfaces[37-38], systematic comparisons of the physical properties of mono-, di-, and trivalent cations inside the EDL remain limited.

Here, with MD simulations, we systematically investigated the ion distribution and ionic solvation of the mono-, di-, and trivalent cations near the negatively charged surface. Under a low surface charge density, all types of cations accumulate in the OHP in a fully hydrated state. As the surface charge density increases, ions in the OHP undergo dehydration to form the IHP. It is worth noting that the order for monovalent cations to migrate from OHP to IHP, $K^+ > Na^+ > Li^+$, is inversely correlated with its static hydration strength order, $Li^+ > Na^+ > K^+$. However, divalent or trivalent cations do not present an IHP in their concentration distributions due to their higher solvation energy. From the analysis of the net charge distribution, the position where cations completely shield the surface charge is closely related to the surface charge density. Charge inversion occurs when the net charge becomes positive, i.e., the positive charges of cations exceed the surface charges, which is more likely to manifest in solutions with divalent or trivalent cations. Our findings indicate that ion distribution, hydration properties, and surface charge screening are strongly influenced by ion size, cation valence, and solution concentration.

## Experimental Section

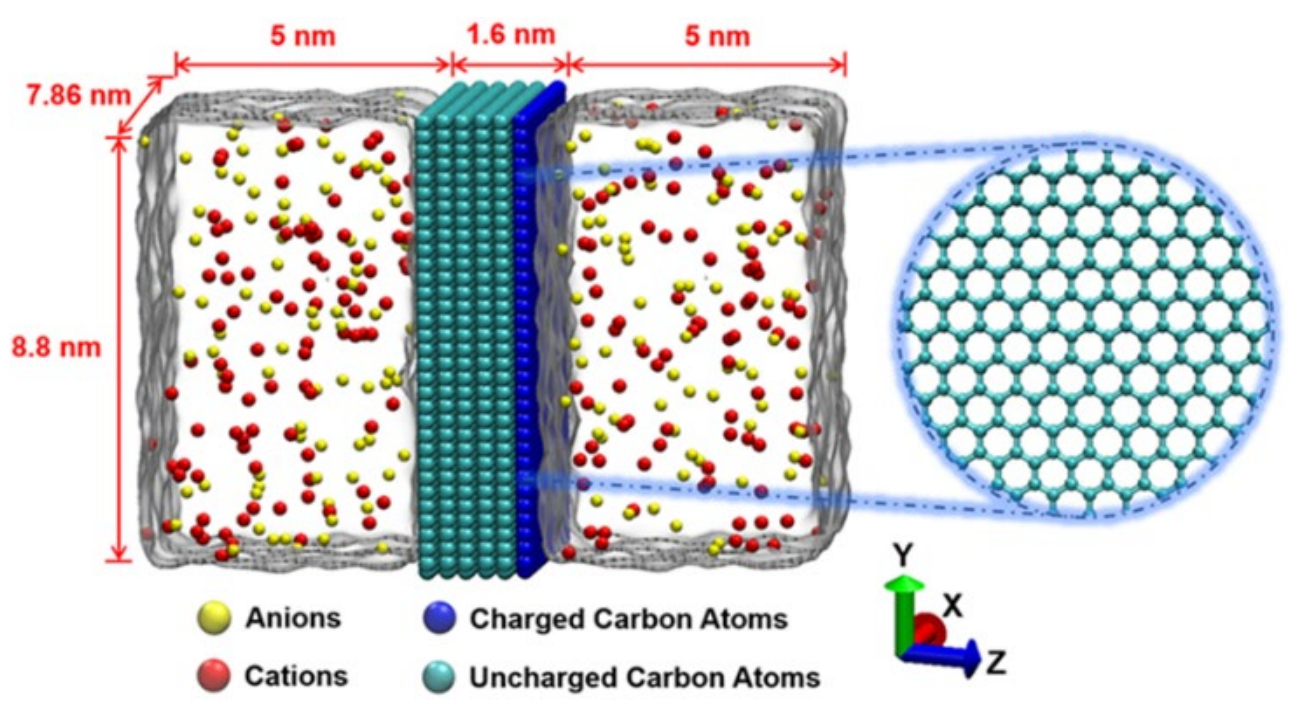

**Figure 1.** Schematic diagram of the MD simulation model. A 6-layer graphene film is located between two aqueous reservoirs. Cyan and blue spheres represent the uncharged and charged carbon atoms. Red and yellow spheres denote the cations and anions in the solution. The zoomed-in part on the right shows the structure of graphene. The whole system is 78.6, 88, and 116 Å in the X, Y, and Z directions, respectively.

To investigate the ion distribution inside the EDLs, VMD 1.9.3[39] was used to construct MD simulation models with a 6-layer graphene film of 78.6 Å in width and 88 Å in height (Figure 1). An offset of 1.4 Å in the Y-axis was set between two neighboring graphene layers (Figure S1).[40-41] Solution reservoirs with a length of ~50 Å were placed on both sides of the graphene film. The simulation model was examined under three different reservoir lengths (Figure S2 and Table S1). From Figure S3, the reservoir size has a negligible effect on the simulation results. The constant charge method[42] was applied in our MD simulations. Carbon atoms of the rightmost graphene layer in the Z direction carried negative charges, whose surface charge density varied from 0 $C/m^2$ to −0.32 $C/m^2$.[43-44] Please note that the graphene is used here to form a solid-liquid interface, without any other special considerations. At different surface charge densities, the charge amounts of individual carbon atoms are listed in Table S2. The mass center of the charged graphene layer was located at the center of the system. Different types of electrolytes, including NaCl, KCl, LiCl, $MgCl_2$, $CaCl_2$, and $LaCl_3$ solutions, were considered in our study. Additional counterions and cations were added to the system to achieve electrical neutrality.[45] The quantities of water molecules and ions in different simulations at various surface charge densities are listed in Table S3.

All simulations were performed with the open-source software Nanoscale Molecular Dynamics (NAMD) 2.13.[46] In our simulation, the TIP3P model[47] was selected to simulate the water molecules. The SETTLE algorithm was chosen to maintain the water molecule geometry.[48] The Lennard-Jones (L-J) potential was used to describe the interactions between related atoms except hydrogen-X pairs (X is an atom species in the solution) and carbon-carbon pairs. The parameters for the L-J potential are listed in Table S4.[49-51] The CHARMM36 force field was chosen to describe the interactions between water molecules, ions, and carbon atoms.[49] Please note that the selected force field neglects polarization effects. Therefore, the dipole response may be underestimated in environments with high electric field gradients or ion-aromatic systems.[52] The cutoff distance and switching distance of the L-J potential were set to 12 Å and 10 Å, respectively.[45, 47] The electrostatic interactions among ions, water molecules, and surface charges were calculated by the Ewald summation algorithm,[53] in which the particle mesh with a 1 Å space grid was employed. Before all the simulations, the velocities of ions and water molecules in the system were randomly distributed following the Maxwell-Boltzmann distribution. Newton's motion equation was integrated by the leap algorithm[54] with a time step of 2.0 fs. During the simulations, the carbon atoms were fixed in their positions. Periodic boundary conditions were applied in all three directions.

For each simulation, the system was initially minimized for 0.1 ns under the NVT ensemble (fixed number of atoms $N$, volume $V$, and temperature $T$). The Langevin thermostat[55] was used to maintain the system temperature constant at 298 K. The system was first energy minimized using steepest descent to eliminate unreasonable configurations, which served as the initial state for simulations.[56] Then, 5 ns was used to equilibrate the system under the NPT ensemble (fixed number of atoms $N$, pressure $P$, and temperature $T$). The pressure was maintained by the Nose-Hoover Langevin method[57] with a period and decay of 0.4 ps and 0.2 ps, respectively. The system equilibrium was determined by calculating the root mean square deviation (RMSD) of the system, which converged after ~4 ns (Figure S4). After equilibration, the system was run for 20 ns under the NVT ensemble, and the last 10 ns were selected for data analysis. The concentration distribution was obtained using a binning method with a 0.1 Å interval. The characteristics of ionic hydration at interfaces were explored by the radial distribution function g(r) and the coordination number (Figure S5).

To provide a comparative analysis, the PB equations (Eqs. 1 and 2) were employed to predict concentration distributions of cations near charged surfaces.

$$\frac{d^2\psi_x}{dx^2} = -\frac{e\rho_0}{\varepsilon_0\varepsilon}\left(z_+ e^{-z_+ e\psi_x/kT} + z_- e^{z_- e\psi_x/kT}\right) \tag{1}$$

$$\rho_x = \rho_0 e^{-ze\psi_x/kT} \tag{2}$$

where $z$, $k$, $e$, $\varepsilon$, $\varepsilon_0$, and $\rho_0$ are the ion valence, Boltzmann constant, elementary charge, dielectric constant of water, dielectric constant of vacuum, and bulk concentration, respectively. The subscripts + and − denote cations and anions. $\rho_x$ and $\psi_x$ are the concentration and electrical potential at the location of x in the perpendicular direction to the surface. The boundary conditions are $\psi_x\big|_{x=0} = \psi_0$ and $\left.\frac{d\psi_x}{dx}\right|_{x=\infty} = 0$, where $\psi_0$ is the electric potential at the charged surface, which can be calculated using the Grahame equation.[1]

**Results and Discussion**

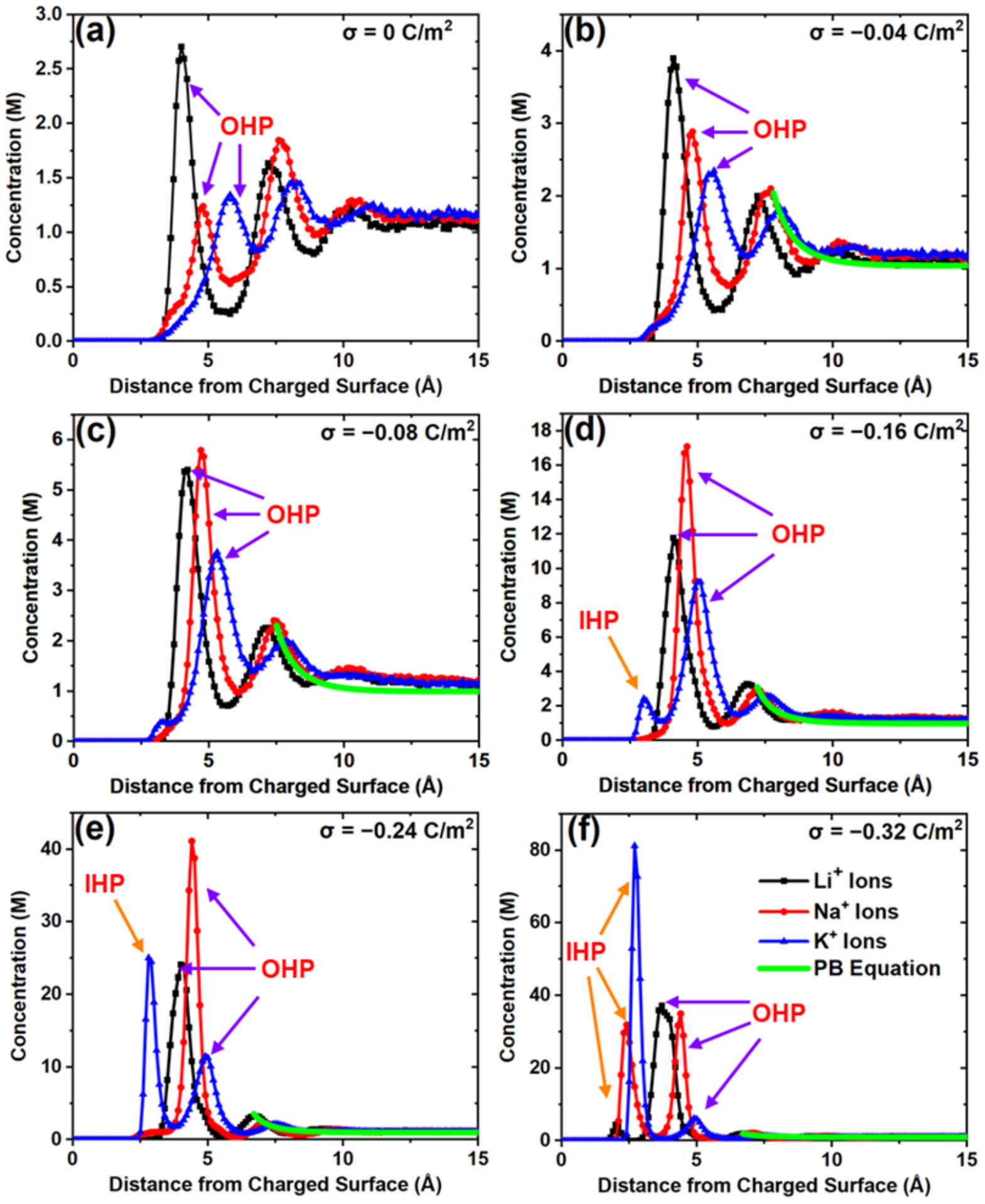

**Figure 2.** Simulated and PB-predicted concentration profiles of monovalent cations in EDLs near differently charged surfaces. (a) 0 C/m$^2$, (b) −0.04 C/m$^2$, (c) −0.08 C/m$^2$, (d) −0.16 C/m$^2$, (e) −0.24 C/m$^2$, and (f) −0.32 C/m$^2$. Three solutions of 1 M were considered, i.e., NaCl, KCl, and LiCl.

In aqueous solutions, the surface charge density has a great influence on the distribution of the ion concentration.[58] Compared with the static approach of the mean-field theory, MD simulations can capture the dynamical concentration distribution and hydration states of ions in the IHP and OHP, which could be affected by ion size, concentration, and surface charge density. Figure 2 shows the concentration distributions of monovalent cations perpendicular to surfaces with different surface charge densities varying from 0 to −0.32 C/m$^2$. In Figure 2a, for neutral surfaces, van der Waals attraction induces the accumulation of both cations and anions, forming an oscillatory distribution within 10 Å from the surface.[59] Cations start to appear at ~2.5 Å from the surface, which can be attributed to the inherent size of cations and surface carbon atoms. $Li^+$, $Na^+$, and $K^+$ ions form the first aggregation peak at distances of ~4.0, ~4.8, and ~5.8 Å from the surface, respectively, corresponding to their hydrated ion sizes.[60] Figure S6 shows that anions begin to occur at ~3.1 Å from the surface. At a low surface charge density of −0.04 C/m$^2$ (Figure 2b), the electrostatic attraction between surface charges and cations motivates the accumulation of cations near the charged surface, which induces the formation of EDLs to screen surface charges effectively. From the distribution of cations, the peak values of the oscillatory profile are enhanced. Cations primarily accumulate inside the OHP, which can be identified by the hydration properties of cations as shown below. Inside EDLs, the steric effect suppresses the peak value for larger cations. Among the three kinds of monovalent cations, $Li^+$ ions have the highest first aggregation peak due to their smallest size.

From Figure 2c, at −0.08 C/m$^2$, more counterions are attracted to the EDLs, showing an increased peak value in the concentration profile. $Na^+$ ions start to display a higher value of the main aggregation peak than $Li^+$ ions. As the surface charge density changes to −0.16 C/m$^2$, the values of the main aggregation peaks are enhanced further. In this case, more $K^+$ ions appear in the IHP. However, no IHP occurs in the concentration profiles of $Li^+$ or $Na^+$ ions. This may be attributed to their higher solvation energy because of their smaller diameters.[18] Please note that whether counterions belong to the IHP or OHP is determined simultaneously by the position and the hydration number. Counterions located in the main aggregation peak and with an equal hydration number to that in the bulk belong to the OHP. And those located at a peak closer to the wall and with a lower hydration number than the bulk value belong to the IHP. For example, the $K^+$ ion peak at ~3.3 Å is identified as the IHP (Figure 2d), because it is closer to the wall, and the average hydration number of $K^+$ ions inside the peak is ~1.5 less than the bulk value.[61] Due to the various sizes of different cations, the IHP and OHP regions are unable to be defined by the constant ion-surface distances.

Figures 2e and 2f show the cases with strongly charged surfaces considered in this work. −0.24 and −0.32 C/m$^2$ are considerable surface charge densities that can be obtained on mica surfaces or surfaces with coatings.[1, 62] At −0.24 C/m$^2$, $K^+$ ions show preferential IHP at ~2.8 Å away from the surface. The accumulation of $K^+$ ions in the IHP results in a pronounced concentration difference between ~25.0 M in the IHP and ~11.5 M in the OHP. At −0.32 C/m$^2$, all three kinds of monovalent cations exhibit an accumulation in the IHP. In LiCl solutions, only a few $Li^+$ ions appear in the IHP. For NaCl, the number of $Na^+$ ions in the IHP is almost the same as that in the OHP. In this case, $K^+$ ions mainly accumulate in the IHP, and the concentration peak reaches ~81.2 M.[63] The IHPs of $Li^+$, $Na^+$, and $K^+$ ions are located at ~2.1, ~2.4, and ~2.7 Å away from the surface, which share the same order as that in OHP, following the increasing order of the ion size. We also considered the cases with 0.5 M in concentration. As shown in Figure S7, the obtained ion concentration distributions present similar characteristics to those at 1.0 M.

The PB equations[9] provide the theoretical predictions of concentration distributions of monovalent cations inside the diffusion layer under moderate surface charge densities varying from −0.04 to −0.16 $C/m^2$. However, as the surface charge density approaches −0.24 $C/m^2$ or a higher value, the predictive accuracy of the PB equations deteriorates due to the enhanced correlation among surface charges and ions. Figure S8 shows the comparison between the PB-predicted monovalent-cation concentration and the simulated concentration profile in the diffusion layer. The corresponding deviation between the prediction and simulation results is plotted in Figure S9.[11, 64-65] Please note that because the surface is negatively charged, a larger surface charge density means a larger absolute value.

Inside EDLs, the accumulation of cations can reduce the surface potential and drive anion clustering near charged surfaces. The monovalent anions and cations distribute alternately within 10 Å of the charged surfaces.[33] The regions beyond 10 Å from the solid surface are considered to be the bulk region, where the concentration value approaches ~1 M. Please note that due to additional counterions added into the system to maintain the electro-neutrality, the bulk concentration of counterions is a little higher than the nominal concentration.

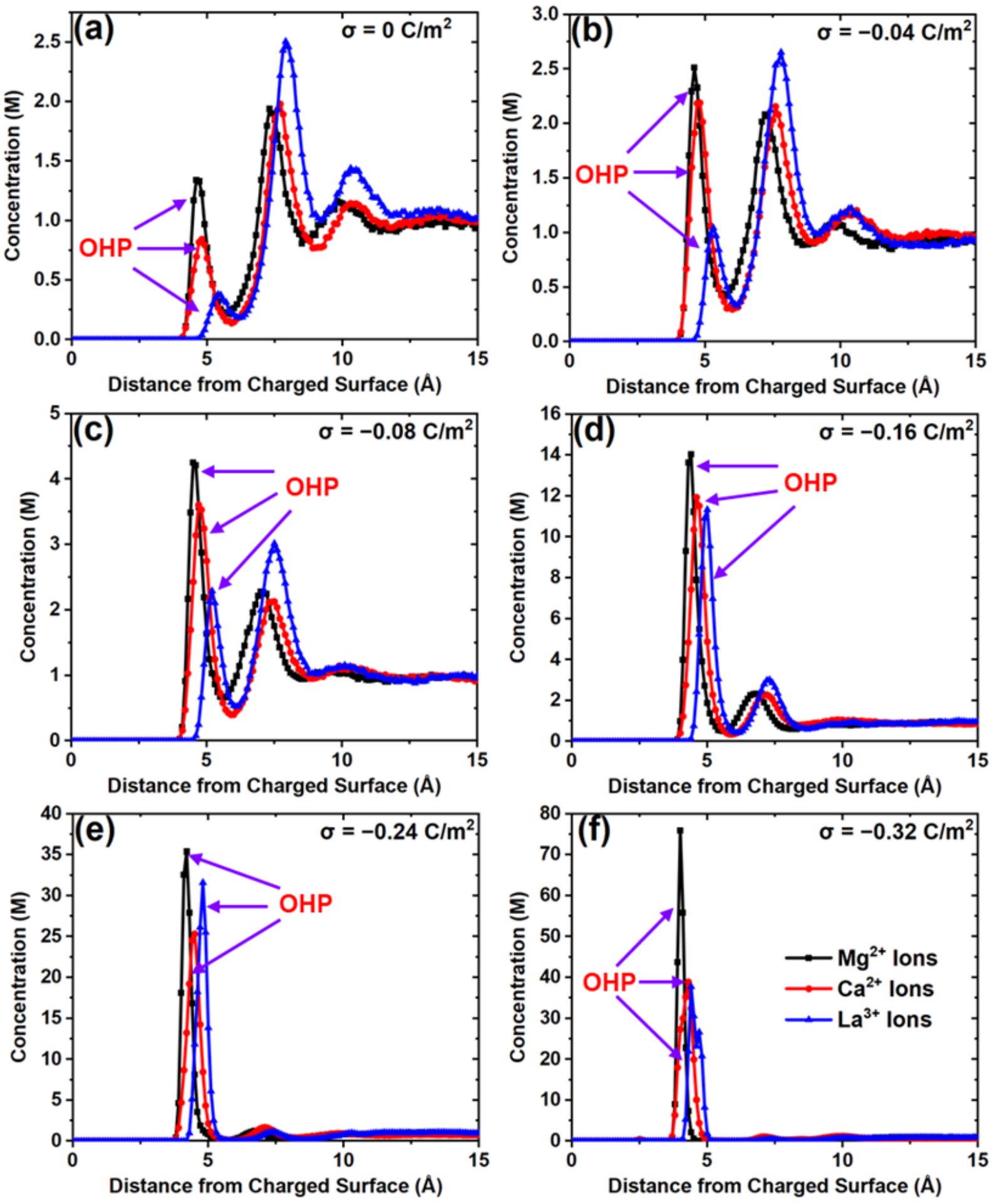

**Figure 3.** Concentration distributions of di- and trivalent cations perpendicular to the negatively charged surfaces with different surface charge densities. (a) 0 C/m$^2$, (b) −0.04 C/m$^2$, (c) −0.08 C/m$^2$, (d) −0.16 C/m$^2$, (e) −0.24 C/m$^2$, and (f) −0.32 C/m$^2$. The cation concentration was kept at 1 M.

Figure 3 shows the concentration distributions of di- and trivalent cations. For solutions with divalent cations, $MgCl_2$ and $CaCl_2$ were considered.[62, 66] $LaCl_3$ solution was employed to introduce trivalent cations. The reliability of the force field parameters for $La^{3+}$ ions was subsequently verified by calculating their diffusion coefficient of ~0.57 × $10^{-9}$ m$^2$/s (Figure S10), which matches well with previously reported experimental results of ~0.62 × $10^{-9}$ $m^2/s$.[67-69] From Figure 3a, $Mg^{2+}$ and $Ca^{2+}$ ions appear at ~4.0 Å from the neutral surface, whereas $La^{3+}$ appears at ~4.8 Å. Figures 3a-3f show a sharp rise in multivalent cation concentrations inside the Stern layer as the surface charge density increases from 0 to −0.32 $C/m^2$. The locations of their accumulation peaks remain at ~5 Å. Although multivalent cations have stronger electrostatic interactions with the charged surface, these ions do not form the IHP even at a high surface charge density.[1, 70] The order of the peak location of the three kinds of cations is $Mg^{2+}$, $Ca^{2+}$, and $La^{3+}$, which corresponds to the order of their ionic radii. This order aligns well with the size-dependent accumulation behavior of monovalent cations observed in Figure 2. Additional simulations were conducted with the concentration of anions setting at 1 M, i.e., 0.5 M $CaCl_2$, 0.5 M $MgCl_2$, and 0.33 M $LaCl_3$. From Figure S11, distributions of cations near charged surfaces share a similar trend. Notably, due to the neglect of the effects of ion interactions and ion size, the PB equations are unable to accurately describe the concentration distribution of multivalent ions.[11, 71]

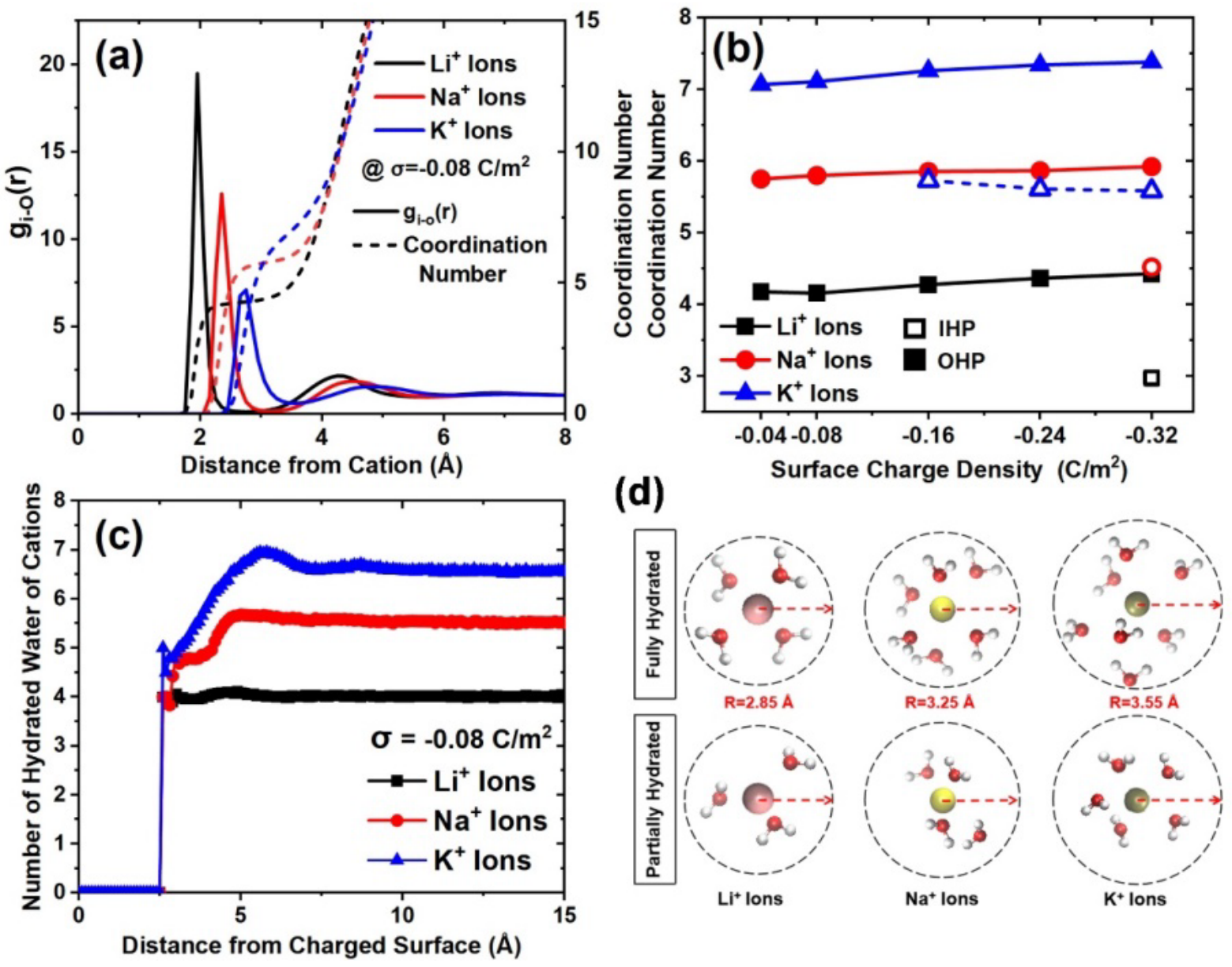

**Figure 4**. Hydration behaviors of monovalent cations inside EDLs. The surface charge density is −0.08 C/m$^2$. (a) Radial distribution functions of oxygen atoms $g_{i\text{-}o}(r)$ around monovalent in the OHP and corresponding coordination number distributions. *i* represents $Li^+$, $Na^+$, and $K^+$ ions, respectively. (b) Coordination number of cations in the IHP and OHP under different surface charge densities. (c) Distribution of hydrated water numbers around cations perpendicular to the surface. (d) Illustration of fully and partially hydrated cations appearing in the OHP and IHP. The dashed circle represents the ionic hydration range, and *R* denotes the hydration radius.

Ionic hydration at solid-liquid interfaces governs the ion distribution and mobility,[72] which is crucial in applications correlated to the hydration lubrication[6] and anti-fouling performance.[73] The hydration behaviors of cations were explored by analyzing the distribution of water molecules near cations. Figure 4a shows the radial distribution functions (RDFs) of oxygen atoms in water molecules around three kinds of monovalent cations in the OHP at −0.08 C/m$^2$. Here, the RDF plotted as a curve $g_{i\text{-}o}(r)$ represents the probability of water molecules appearing at the location *r* around the cations.[74] The ion radius governs the electrostatic interactions between cations and water molecules. Smaller cations show higher RDF peaks and peak locations closer to the cations, indicating a stronger hydration property. The position of the first valley in the RDF profile represents the boundary of the first hydration shell, which is generally considered as the ionic hydration radius.[75] Due to the smallest size, $Li^+$ ions have the largest first peak value in the RDF profile, corresponding to the strongest solvation energy in the three monovalent cations. Figure S12 shows the RDF profiles of di- and trivalent cations. The hydration radii of $Li^+$, $Na^+$, $K^+$, $Mg^{2+}$, $Ca^{2+}$, and $La^{3+}$ ions are calculated as 2.9, 3.1, 3.6, 3.0, 3.4, and 3.6 Å, respectively. These values align with those in the bulk case, indicating that the surface charges do not affect the intrinsic hydration properties of the cations (Figure S13).[70] The coordination numbers, as shown by the dashed line in Figure 4a, are derived by integrating the number of water molecules in the radial direction. This profile quantifies the average number of water molecules in the first hydration shell surrounding the cation. For $Li^+$, $Na^+$, and $K^+$ ions, the respective coordination numbers are ~4.2, ~5.8, and ~7.1 in the OHP. The coordination numbers of $Mg^{2+}$, $Ca^{2+}$, and $La^{3+}$ ions are ~5.6, ~7.3, and ~7.6, respectively (Figure S12).

Figure 4b summarizes the coordination numbers of the three monovalent cations inside the EDLs at various surface charge densities. As the surface charge density increases from −0.04 to −0.32 $C/m^2$, the coordination numbers of cations in OHP remain nearly constant, i.e., ~4.2, ~5.8, and ~7.1 for $Li^+$, $Na^+$, and $K^+$ ions. Figure 4d compares the solvation structures of the three cations inside the OHP and IHP. When the electrostatic interaction between cations and surface charges is weaker than the solvation energy, the cation remains distant from the surface and maintains a fully hydrated state. As the surface charge density increases or cations approach the charged surface gradually, the enhanced electrostatic attraction overcomes the ionic solvation energy, which can strip off part of the hydrated water molecules around cations (Figure S14).[76] The dehydrated cations can continue to approach the charged surface to form an IHP, while the fully hydrated cations form OHP. This dehydration process of cations from the OHP to IHP is affected by their hydration radius. $K^+$ ions with the largest hydration radius form the IHP at −0.08 C/m$^2$, while $Li^+$ ions with the smallest hydration radius only dehydrate to form the IHP at −0.32 C/m$^2$. To reach the IHP from the OHP, $K^+$, $Li^+$, and $Na^+$ ions need to dehydrate ~1.5, ~1.5, and ~1.4 water molecules, respectively.

Figure 4c exhibits the distribution of ionic hydration numbers perpendicular to the surface at −0.08 C/m$^2$. Cations begin to appear at ~2.5 Å from the surface, where their hydration number is relatively low. Due to its small ion radius and absence of the IHP, the hydration number of $Li^+$ ions remains stable at ~4. From the captured scenarios (Figure 4d), $Li^+$ ions can form a relatively stable tetrahedral structure with four water molecules around them.[18, 77] However, $Na^+$ and $K^+$ ions exhibit a disordered solvation structure (Figure S15), due to their lower solvation energy than that of $Li^+$ ions. The hydration number of $Na^+$ ions increases from ~4 to ~5.6 in the region from ~2.5 to ~5.0 Å beyond the surface. The hydration number of $K^+$ ions increases rapidly to ~6.9 within a short distance from ~2.5 to ~5.7 Å, then has a slight decrease to the bulk value of ~6.5 due to the gradually increased anions and cations. $Mg^{2+}$, $Ca^{2+}$, and $La^{3+}$ ions present no dehydration, and their hydration number remain stable at ~5.6, ~7.3, and ~7.6, respectively.

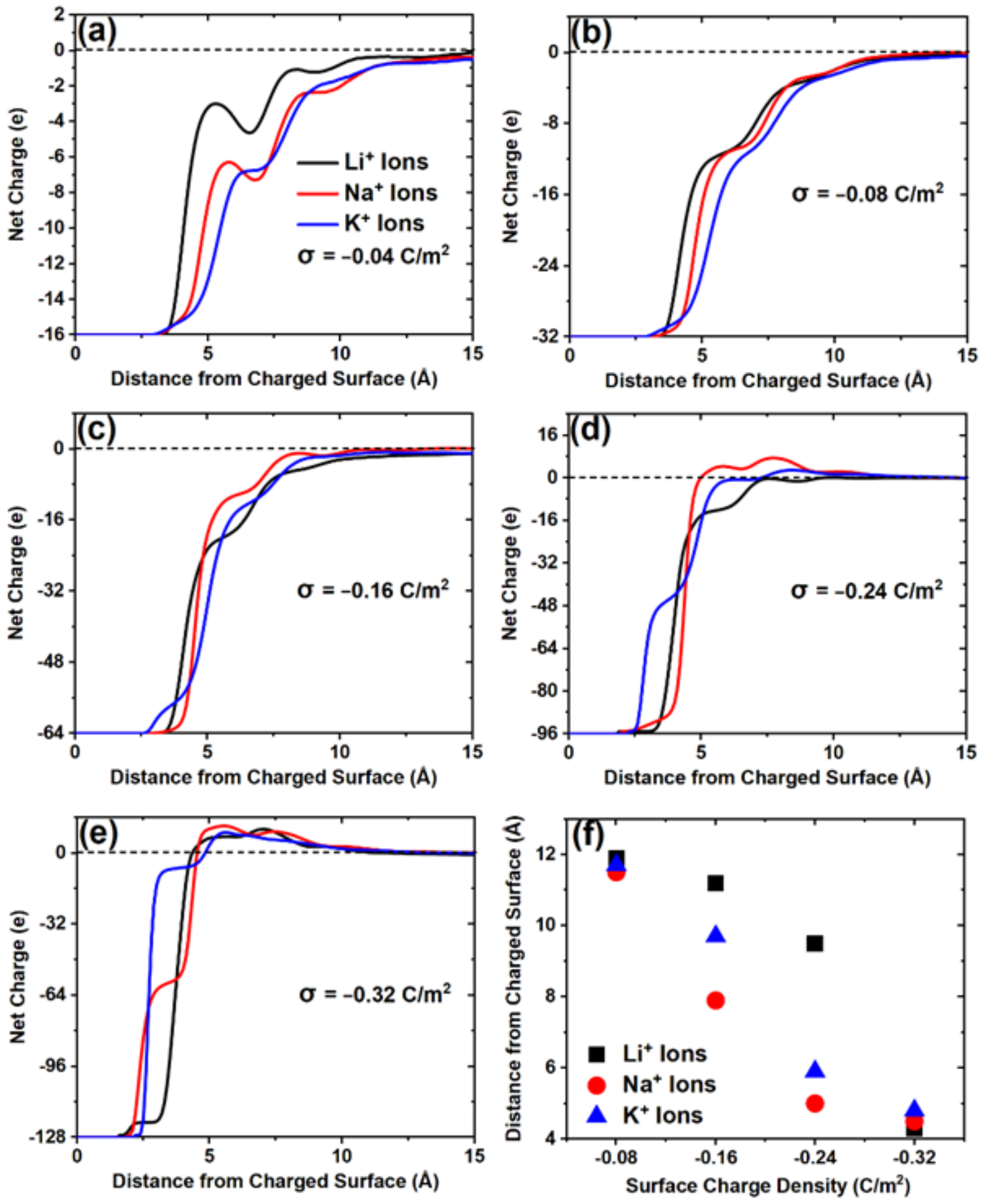

**Figure 5.** Screening behaviors of the three monovalent cations to surface charges. (a-e) Distributions of the net charge perpendicular to solid surfaces at different surface charge densities. (a) −0.04 C/m$^2$, (b) −0.08 C/m$^2$, (c) −0.16 C/m$^2$, (d) −0.24 C/m$^2$, and (e) −0.32 C/m$^2$, (f) Location of the neutral plane at different surface charge densities.

The shielding effect of cations on surface charges can be characterized based on the ion distributions.[33, 78] Figure 5 shows the distributions of the net charge perpendicular to surfaces under various surface charge densities. At each position, the net charge was calculated by integrating the local charge of ions and surface charges ($Q_s$) by Eq. 3.

$$Q_{\mathrm{net}}(x) = Q_s + \sum_{i=1}^{M} \left[ N_+(i) \times z_+ + N_-(i) \times z_- \right] \tag{3}$$

where $Q_{net}(x)$ represents the net charge at a distance $x$ from the charged surface, $Q_s$ is the surface charge number, $M = \mathrm{int}(x/0.1)$. $i$ is the bin number, $N_+(i)$ and $N_-(i)$ are the local numbers of cations and anions at position $I$, $z_+$ and $z_-$ are the valences of cations and anions.

As shown in Figure 5a, at −0.04 C/m$^2$, the counterions are attracted to accumulate in EDLs to screen the surface charges. In this case, the net charge distributions in the three monovalent solutions share a similar trend. $Li^+$ ions present a more effective screening to surface charges within 10 Å away from the surface due to the highest concentration peak in EDLs (Figure 2b). At the position of ~12 Å from the surface, the surface charges are completely screened, denoting a neutral plane. In the cases at −0.08 and −0.16 C/m$^2$, due to increased counterion accumulation in EDLs, the location of the neutral plane gradually moves closer to the charged surface. Simultaneously, $Na^+$ ions start to exhibit a better screening performance among the three monovalent ions. This arises from the increased accumulation of $Na^+$ ions in the range of 5 to 10 Å near the surface, as seen from the ion concentration distributions in Figures 2c-d.

As the surface charge density increases to −0.24 C/m$^2$, $K^+$ ions in the IHP can screen half of the surface charges while the remaining unscreened surface charges continue attracting additional $K^+$ ions to form the OHP. At ~7.5 Å, the net charge becomes positive due to the excess counterions accumulating inside EDLs, representing the appearance of the charge inversion.[11] This phenomenon has been observed in many studies that can not be predicted by the mean-field theory. The strong correlation theory attributes it to the correlation among ions and surface charges, which positively correlated to the surface charge density, ion valence, and salt concentration. The non-electrostatic interactions between ions and surfaces, such as van der Waals forces, have no obvious promoting effect on charge inversion.[24] The charge inversion reported here is derived from the steady-state ion concentration distribution, reflecting the average ionic behavior across the entire solid-liquid interface rather than local or transient overcompensation. In previous simulation and experimental investigations,[79-81] charge inversion may occur at the solid-liquid interfaces in monovalent solutions facilitated by the strong surface charge and high electrolyte concentration.[11] For $Na^+$ ions, the neutral plane occurs at a position of ~5 Å from the surface. Beyond this position, in the range of ~5.0 to 12.2 Å from the charged surface, excessive accumulation of $Na^+$ ions in EDLs triggers the charge inversion. With the surface charge density increasing further to −0.32 C/m$^2$, all three cations fully shield the surface charges at ~4.7 Å from the surface. Near the strongly charged surfaces, the charge inversion happens in all three solutions with comparable positive net charges.

Figure 5f exhibits the variation of the location of neutral planes with the surface charge density. For all three kinds of monovalent cations, a similar decreasing trend is observed. With the surface charge density varying from −0.04 to −0.32 C/m$^2$, the location of the neutral plane changes from ~12 to ~4 Å away from the surface (Table S5).

In simulations with three kinds of monovalent cations, we observed that charge inversion appears with $K^+$ and $Na^+$ ions at a charge density of −0.24 C/m$^2$. This is because the larger ionic radius and lower solvation energy of $K^+$ and $Na^+$ ions lead to a higher accumulation within the EDLs. This accumulation enhances the ion-ion correlation, which induces the charge inversion. In contrast, $Li^+$ ions present the charge inversion only at −0.32 C/m$^2$. Our results align well with that ion size promotes the correlation among ions in EDLs. Smaller ions like $Li^+$ ions, due to their stronger hydration shell, are more restricted in their approach to the surface, delaying charge inversion. Solution concentration also influences the ion distribution in EDLs (Figure S16). In solutions with a higher concentration, the plane where the surface charge is completely screened becomes closer to the surface. As predicted by the strong correlation theory, a high salt concentration enhances the correlation among ions and surface charges at solid-liquid interfaces, thereby promoting the charge inversion.[82] In Figure S16, at −0.16 C/m$^2$ and the concentration of 2.0 M, all three types of salts exhibit charge inversion. From our results, for monovalent solutions, the degree of charge inversion is directly related to the surface charge density, ion size, and salt concentration.

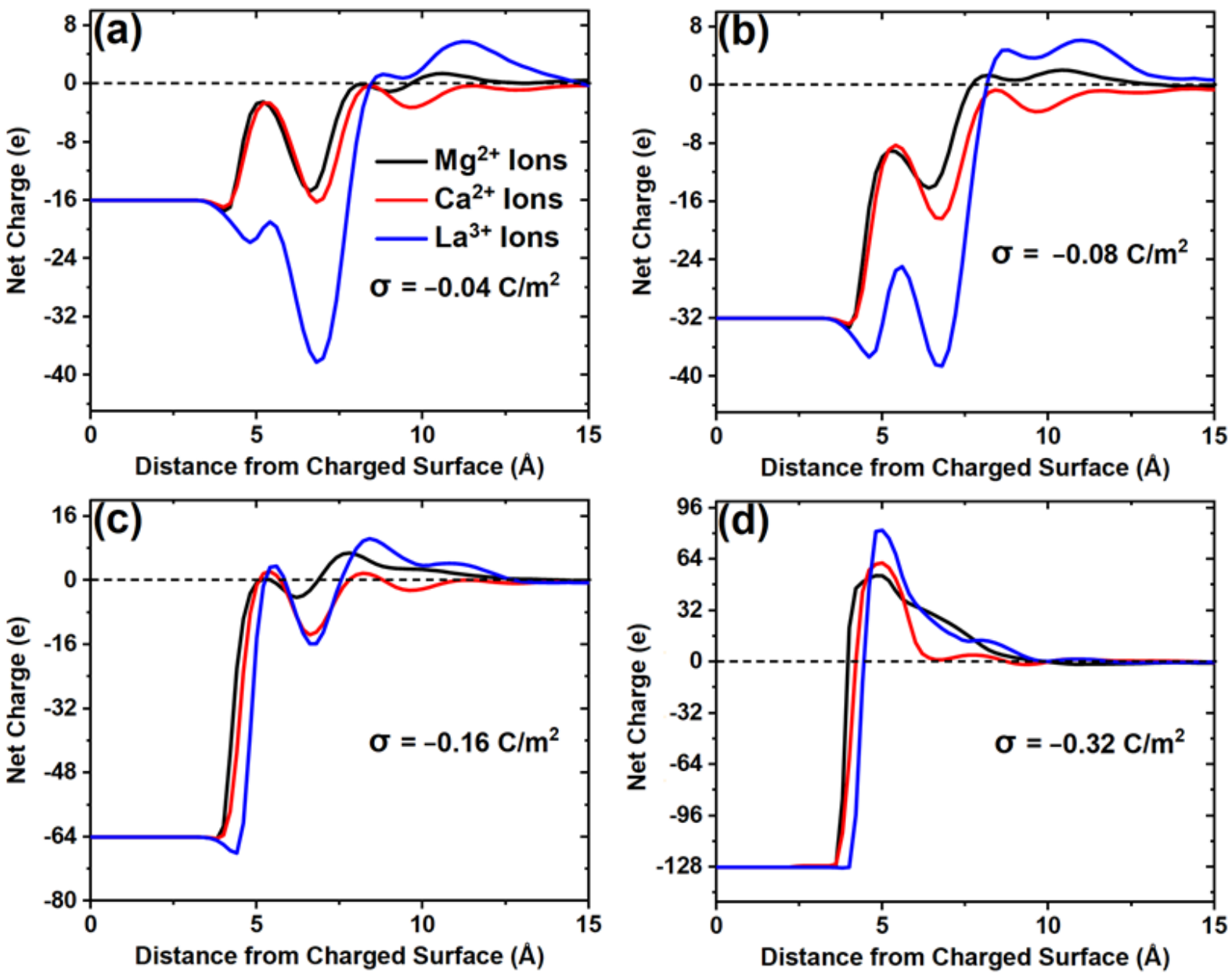

**Figure 6.** Screening behaviors of the divalent and trivalent cations to the surface charges at different surface charge densities. (a) −0.04 C/m$^2$, (b) −0.08 C/m$^2$, (c) −0.16 C/m$^2$, and (d) −0.32 C/m$^2$. The cation concentration was set to 1 M.

Figure 6 shows the screening behavior of divalent and trivalent cations on the surface charges. Compared with monovalent cations, divalent cations have a stronger shielding effect on surface charges. At −0.04 and −0.08 C/m$^2$, the net charge exhibits an oscillatory distribution between ~3.4 to ~8.4 Å from the charged surfaces. This can be attributed to the alternating distribution of cations and anions within 10 Å away from the charged surface. In electrolytes with multivalent cations, the oscillation in the distribution of net charges becomes more significant due to the greater charge of accumulated ions. In Figure 2d, at −0.16 C/m$^2$, a slight charge inversion emerges, which intensifies progressively as the surface charge density increases to −0.32 C/m$^2$. However, in the $LaCl_3$ solution, the charge inversion can occur at −0.04 C/m$^2$. This indicates that multivalent cations exhibit stronger electrostatic interactions with surface charges, enabling the appearance of charge inversion at lower surface charge densities, consistent with the strong correlation theory.

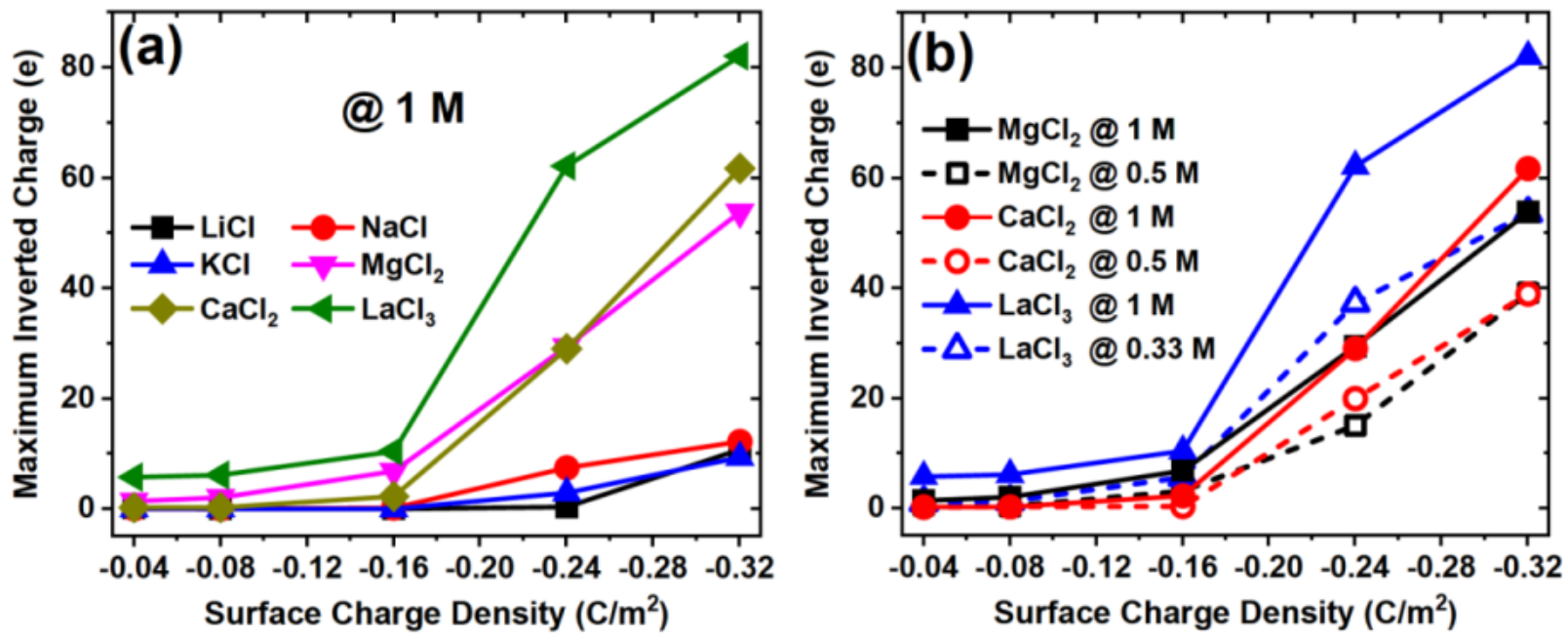

**Figure 7.** Behaviors of charge inversion in solutions with the mono-, di-, and trivalent cations at different surface charge densities. (a) Maximum inverted charges in solutions with different cations of 1 M, (b) Maximum inverted charges in solutions with multivalent cations of different concentrations.

To further investigate the relationship between the ion valence and screening efficacy to surface charges, we analyzed the maximum inverted charge under various surface charge densities, shown in Figure 7. In LiCl solution, the charge inversion only occurs at −0.32 C/m$^2$, with a peak positive charge is ~9.3 e. For $Na^+$ and $K^+$ ion solutions, the charge inversion appears at −0.24 C/m$^2$, with maximum inverted charges of ~7.5 e and ~2.9 e, which rise to ~12.2 e and ~9.3 e at −0.32 C/m$^2$, respectively. In divalent solutions, $Mg^{2+}$ and $Ca^{2+}$ ions start to exhibit charge inversion at −0.04 and −0.16 C/m$^2$, respectively, with the maximum inverted charges of ~1.4 e and ~2.3 e, which dramatically increase to ~53.8 e and ~61.7 e at −0.32 C/m$^2$. For $La^{3+}$ ions, the charge inversion occurs at −0.04 C/m$^2$, and the maximum inverted charge reaches ~82.1 e at −0.32 C/m$^2$, corresponding to 64% of the surface charge quantity (Figure S17). As the cation valence increases, the charge inversion occurs at a lower surface charge density and exhibits a larger value of positive net charge. In Figure 7b, we compared the influence of the salt concentration on the charge inversion. As the concentration of multivalent cations in the solution decreases, the amount of inversed charges reduces. For example, under a surface charge density of −0.32 C/m$^2$, the maximum inverted charge decreases by ~28 e in 0.33 M $LaCl_3$ solutions, corresponding to a ~22% reduction in the maximum inverted charge at 1 M (Figure S17). Our results exhibit that charge inversion is modulated simultaneously by the ion valence, size, concentration, and surface charge density.

The investigation of charge inversion at solid-liquid interfaces may have potential applications in biomedical engineering and energy storage. Charge inversion phenomena have been observed in biological systems, such as the condensation of DNA or protein molecules by multivalent cations, suggesting a potential regulatory role in biomolecular assembly and function.[83] In multivalent solutions, the charge inversion of DNA molecules may enable gene therapy across negatively charged cell membranes.[84] In highly concentrated electrolyte solutions of supercapacitors, the exploration of the mechanism for charge inversion at the electrode-liquid interface is crucial for optimizing the performance of the energy storage device, because charge inversion affects the ion structures in porous electrodes.[85-86]

## Conclusions

Through MD simulations, we investigated the ion distributions, hydration numbers, and screening performance of cations near charged surfaces. At low surface charge densities, the cations accumulate only in the OHP. As the surface charge density increases, the stronger electrostatic attraction pulls counterions closer to the surface. When this attraction is sufficient to disrupt their hydration shells, counterions undergo dehydration to form the IHP. $Li^+$ ions, with the smallest ion size among the three monovalent cations considered, are positioned closest to the charged surface. At −0.08 C/m$^2$, $K^+$ ions begin to aggregate at ~3.3 Å from the charged surface, implying the formation of an IHP. In contrast, $Na^+$ and $Li^+$ ions start to exhibit IHP at a higher surface charge density of −0.24 and −0.32 C/m$^2$. Notably, the coordination number of cations in the OHP remains constant under varying surface charge densities from 0 to −0.32 C/m$^2$. While cations in the IHP consistently exhibit a coordination number ~1.5 lower than that in the OHP, indicating that cations in the OHP need to lose ~1.5 water molecules to reach the IHP. From distributions of the net charge perpendicular to the surfaces, as the surface charge density increases from −0.08 to −0.32 C/m$^2$, the location of the neutral planes gradually decreases from ~11.7 to ~4.5 Å. In addition, excess counterions accumulated in EDLs can induce charge inversion, which degree is closely related to the ion size, surface charge density, and solution concentration. Our results provide theoretical insights into the factors governing ion distributions and hydration characteristics at solid-liquid interfaces, offering guidance for optimizing interfacial properties in applications such as desalination systems and biomolecule detection platforms.

**Supporting Information**

See supplementary material for additional simulation details, including the number of atoms in the system, L-J potential parameters, and a schematic diagram of the graphene structure; calculation methods, including the calculation method of the radial distribution function; and additional simulation results, including the ion and water concentration profiles at varying charge densities, deviation between PB-predicted ion concentration and MD results at different surface charge densities, and the screening behavior of the ions.

**Acknowledgment**

The authors thank Prof. Mamadsho Ilolov at the Center of Innovative Development of Science and New Technologies (National Academy of Sciences of Tajikistan) for carefully reading and improving the manuscript. This research was supported by the National Key R&D Program of China (2023YFF0717105), the Shenzhen Science and Technology Program (JCYJ20240813101159005), the Guangdong Basic and Applied Basic Research Foundation (2023A1515012931 and 2025A1515010126), the Shandong Provincial Natural Science Foundation (ZR2024ME176), the National Natural Science Foundation of China (52105579), the Ningbo Yuyao City science and technology plan project, China (2023J03010010), and the Innovation Capability Enhancement Project of Technology-based Small and Medium-sized Enterprises of Shandong Province (2024TSGC0866),.

**TOC**

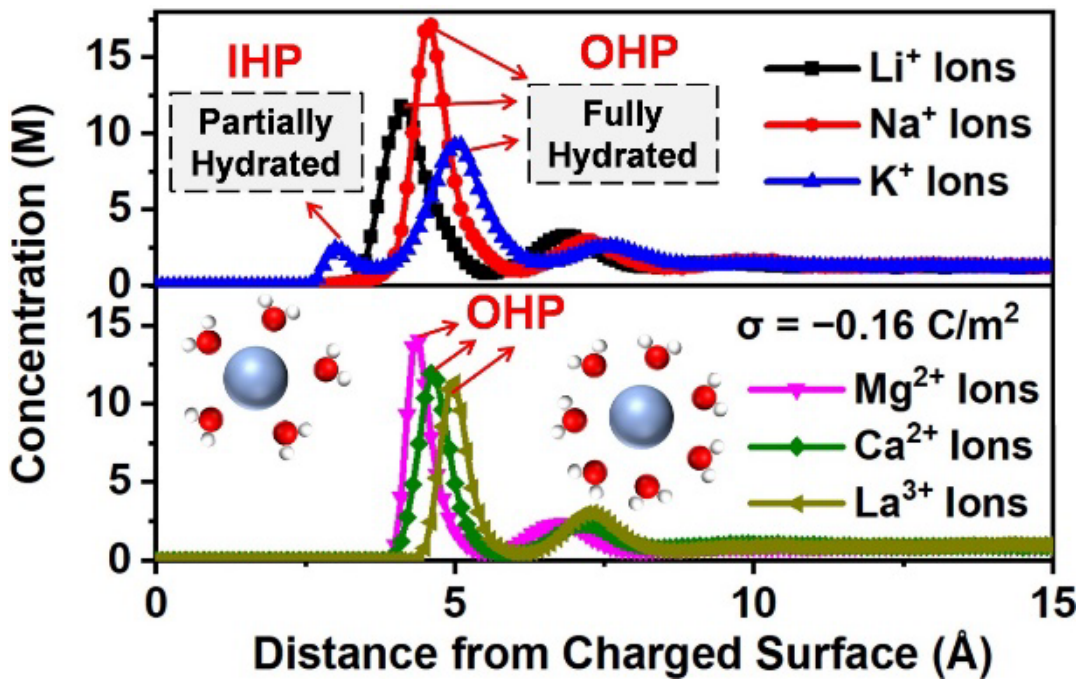